 \documentclass[11pt,twoside]{article}
 \usepackage{Cargesepasp,epsf}
 \newcommand{\markcite}[1]{}
 \newcommand{\Msun}{M_\odot}
 \def\lesssim{\lower.5ex\hbox{$\; \buildrel < \over \sim \;$}}
 \def\gtsim{\lower.5ex\hbox{$\; \buildrel > \over \sim \;$}}

\newcommand{\vesc}{v_{\rm esc}}
\newcommand{\Mej}{M_{\rm ej}}
\begin{document}
\title{Feedback from Protostellar Outflows in Star and
Star Cluster Formation}
\author{Christopher D. Matzner}
\affil{CITA, McLennan Labs, 60 St. George St, Toronto, Ontario M5S
3H8, Canada}

\begin{abstract}
Magnetic stresses collimate protostellar winds into a common
distribution of force with angle. Sweeping into the ambient medium,
such winds drive bipolar molecular outflows whose properties are
insensitive to the distribution of ambient gas and to the details of
how the wind is launched, and how its intensity varies over
time. Moreover, these properties are in accord with the commonly observed
features of outflows. 

This model is simple enough to permit a quantitative study of the
feedback effects from low-mass star formation. It predicts the rate at
which star-forming gas is ejected by winds, and hence the efficiency
with which stars form. Applied to individual star formation, it relates
the stellar initial mass function  to the distribution of pre-stellar
cores. Applied to cluster formation, it indicates whether the
resulting stellar system will remain gravitationally bound. 

Using the energy injection and mass ejection implied by this model, we
investigate the dynamical evolution of a molecular clump as a stellar
cluster forms within it. This depends critically on the rate at which
turbulence decays: it may involve equilibrium star formation (slow
decay), overstable oscillations, or collapse (fast decay). 
\end{abstract}

\section{Introduction}
Stars, even low-mass stars, are not born quietly. In the process of
accreting gas through an accretion disk, each star returns a fraction
of the inspiralling material to the interstellar medium at speeds
comparable to the escape velocity from the stellar surface. These
powerful winds, observable as jets of ionized gas, sweep the ambient
material into opposed streams known as bipolar molecular
outflows. Because of their intensity, protostellar winds pose severe
challenges to those who wish to model star forming regions.  On a
small scale, the wind from a forming star is capable of stripping gas
from its parent molecular core even as that core feeds material to the
star. On a larger scale, the outflow driven by a star's wind can punch
through a massive molecular clump that is forming a stellar
cluster. There is now plentiful observational evidence that these
interactions are common, even ubiquitous, within star-forming
clouds. Moreover, there is evidence to support the argument
\markcite{NS80}({Norman} \& {Silk} 1980) that the feedback from
protostellar winds is the primary driver of turbulence within
molecular clouds, and may be responsible for establishing Larson's
(\markcite{1981MNRAS.194..809L}1981) line width-size relation in
these clouds \markcite{M89}({McKee} 1989).

Because these phenomena have profound consequences for the formation
of stars and star clusters and for the evolution of molecular clouds,
it is important to quantify the ways in which protostellar winds
impact their molecular environment. In the past, models of feedback in
star formation have either ignored the effects of low-mass stars
\markcite{1983MNRAS.203.1011E}(e.g., {Elmegreen} 1983) or considered
only spherically symmetric protostellar winds
\markcite{NS80,1995ApJ...450..183N}(e.g., {Nakano}, {Hasegawa}, \&
{Norman} 1995), despite the fact that these winds are clearly
jet-like. More sophisticated models of protostellar outflows
\markcite{Shuea91,MC93,1996ApJ...472..211L}(such as those of {Shu}
{et~al.} 1991; {Masson} \& {Chernin} 1993; {Li} \& {Shu} 1996) have
not been applied to the basic questions of feedback. Partly, this has
been caused by a lack of consensus \markcite{Richeretal}(see {Richer}
{et~al.} 2000) as to which of the several classes of models for
outflows best describes their dynamics.

We shall argue that there is a single model for protostellar outflows
that is both motivated by theories of protostellar winds and
validated by its ability to explain some of the common, otherwise
mysterious features of protostellar outflows. Although this model
cannot explain all the details of these flows, it provides a stable
platform from which to launch an exploration wind-cloud
interactions. We shall then use this model to address basic questions
of star formation: the stellar initial mass function (IMF); the
efficiency of star formation in clusters; and the dynamical evolution
of star-forming gas during stellar cluster formation.

The work presented here has been conducted in collaboration with Chris
McKee and appears or is intended to appear in:
\markcite{myphd}{Matzner} (1999),
\markcite{1999ApJ...526L.109M}{Matzner} \& {McKee} (1999a),
\markcite{effic}{Matzner} \& {McKee} (2000), and
\markcite{MBM2000}{Matzner}, {Bertoldi}, \& {McKee} (2000).

\section{Common Properties of Bipolar Molecular Outflows}
\label{S:outflowproperties}  
Protostellar molecular outflows typically share several common
features, as discussed by \markcite{MC93}{Masson} \& {Chernin} (1993)
and \markcite{LF96}{Lada} \& {Fich} (1996). These are: a roughly
linear position-velocity (PV) diagram (a ``Hubble law''); a lack of
receding material in the approaching lobe, and vice versa; and a power
law distribution of mass with velocity: $dM/v_{\rm obs}\propto v_{\rm
obs}^\Gamma$, where $v_{\rm obs}$ is the line-of-sight velocity
relative to the systemic velocity and $\Gamma$ is typically about
-1.8. Of these, the power law distribution of mass with velocity is
the most difficult to explain, as it often holds quite well over a
factor of five or ten in $v_{\rm obs}$ \markcite{LF96}(e.g., {Lada} \&
{Fich} 1996).

In order to construct a theory for the feedback effects of
protostellar winds it is necessary to develop a model for these
outflows, which represent the interaction between winds and the
ambient gas. To be tenable, such a model must confront the common
properties of outflows listed above. Moreover, it must also be
compatible with the ways in which protostellar winds are launched and
collimated, a topic we now address.

\section{Structure and Intensity of Protostellar Winds}\label{S:WindStructure}
Protostellar winds are launched centrifugally from the inner regions
of protostellar accretion disks
\markcite{1982MNRAS.199..883B}({Blandford} \& {Payne} 1982). This is
made possible by the presence of intense poloidal magnetic fields,
which act as rigid tubes anchored in the disk, along which gas is
flung away. Once the wind has traveled sufficiently far, the field
weakens and gas begins to travel ballistically rather than being
forced to corotate with the disk.
Since it is still stuck to the same field line as when it was
launched, and since that field line stretches back to the rotating
disk, the magnetic field is wound into a tight spiral. 
By the Biot-Savart law, the tightly-wound field implies that the wind
encloses a current.
Magnetic stresses associated with current within the wind can generate
forces that push the wind toward or away from its axis. 
These forces are important for the collimation of the wind on scales
($\lesssim 1$ AU) comparable to the launching region of the accretion
disk, but on the larger scales of molecular outflows ($\sim 0.1$ pc),
the wind relaxes into transverse pressure balance. 
Therefore, the wind's magnetic flux lines tend toward a force-free
state in which there is no current within the wind: the enclosed
current lies entirely along the axis, rather than threading the wind
itself. The field thus settles into the force-free state $B_\phi =
2I(r)/(c\varpi)$, where $\varpi= r\sin \theta$ is the cylindrical
radius and the current $I(r)$ is a very slowly varying function of
$r$. (There must be currents, and hence forces, in the wind for $I(r)$
to vary at all: these effect a continuing gradual collimation of the wind.)

This force-free distribution of toroidal magnetic field is sufficient
to determine the ram pressure of the wind at large distances from its
source. For, the magnetic field lines must corotate with the disk in a
steady state. This, along with the fact that the wind will expand to
fill the entire solid angle available to it, allows one to trace the
field lines from their origin on the disk to their destination on the
sky \markcite{1997ApJ...486..291O,myphd}({Ostriker} 1997; {Matzner}
1999). Then, the conservation of mass and energy along each streamline
determines the ram pressure of the wind in each direction. In the case
of a wind whose density just above the disk varies with disk radius
$\varpi_0$ as $\rho_{w0}\propto \varpi_0^{-q}$, and whose field lines
expand significantly ($\varpi\gg\varpi_0$) according to the force-free
distribution of the field, the wind ram pressure varies as $\rho_w
v_w^2 \propto C(r) \varpi_0^{(1-q)/2} \varpi^{-2}$
\markcite{1997ApJ...486..291O}({Ostriker} 1997).  Now, the disk radius
$\varpi_0$ is generally restricted within a very narrow range compared
to the wind radius $\varpi$: for instance, in the model of
\markcite{1995ApJ...455L.155S}{Shu} {et~al.} (1995), $\varpi_0$ takes
only a single value whereas $\varpi$ extends from a few AU to $\sim
0.1$ pc. For this reason, the ram pressure of the wind is very close
to $\rho_w v_w\propto \varpi^{-2}$ in practice
\markcite{1999ApJ...526L.109M}({Matzner} \& {McKee} 1999a). This force
distribution, which is characteristic of the x-wind model
\markcite{1995ApJ...455L.155S}({Shu} {et~al.} 1995), is thus common
among hydromagnetic winds.

In reality, variability and instabilities are likely to broaden the
wind force inside some angle $\theta_0$; therefore, we consider the
smoothed distribution
\begin{equation} \label{eq:windforce}
\rho_w v_w =\frac{\dot{m}_w v_w}{4\pi r^2} \frac{1}{\ln(2/\theta_0)}
\frac{1}{(\sin\theta)^2 + (\sin \theta_0)^2}. 
\end{equation} 
In the model for outflows presented below, observations imply
$\theta_0\simeq 0.01$.

The fact that hydromagnetic winds tend to $\rho_w v_w\propto
(\sin\theta)^{-2}$ is consistent with the theoretical expectation
(Matzner \& McKee 1999a) that the fast Alfv\'enic Mach number,
$B^2/(4\pi \rho_w v_w^2)$, is roughly constant. 
This expectation follows from the fact that this ratio is unity near
the source, and varies logarithmically with distance along each
streamline. 
The $(\sin\theta)^{-2}$ force distribution implies that the cumulative
wind momentum within some angle, $p_w(<\theta) \equiv \int^\theta 2\pi
r^2 \rho_w v_w \sin\theta' d\theta'dt$, varies as
$\ln(\theta/\theta_0)$ for a broad range of angles, and this property
will be useful for understanding outflows and their effects.

A protostar's wind lasts as long as it accretes material through its
disk, which is essentially the free-fall time of the unstable core
from which it formed, typically $\sim 10^5$~yr. If the wind's mass is
a fraction $f_w$ of the star's mass, then the wind momentum is $f_w
v_w m_\star$.  We estimate that $f_w v_w \simeq 40$~km~s$^{-1}$,
roughly consistent with the observational estimate of Richer et
al. (2000); for a wind velocity $v_w\simeq 200$~km~s$^{-1}$, this
implies $f_w\simeq 1/5$, between the theoretical predictions of Shu et
al. (1988) and Pelletier and Pudritz (1992). We expect $f_w v_w$ to be
approximately independent of stellar mass, because $v_w$ scales
roughly with the stellar escape velocity, which itself is regulated by
deuterium burning during accretion
\markcite{1988ApJ...332..804S}({Stahler} 1988).

\section{Bipolar Molecular Outflows from Hydromagnetic Protostellar
Winds} \label{S:Outflows} 

\begin{figure}
\plotone{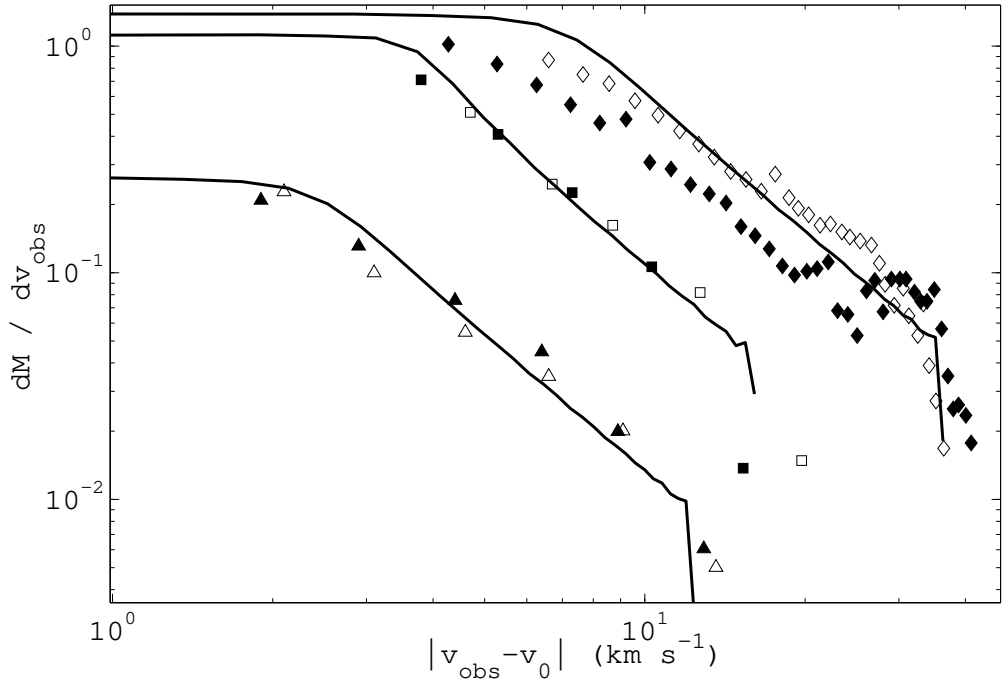}
\caption{\small Mass-velocity relations for L1551 ({\em triangles},
\markcite{MS88}{Moriarty-Schieven} \& {Snell} 1988), NGC2071 ({\em
squares}: \markcite{MSH89}{Moriarty-Schieven}, {Hughes}, \& {Snell}
1989), and NGC2264G ({\em diamonds}: \markcite{LF96}{Lada} \& {Fich}
1996), fit with outflow model {\em solid lines}). The
inclination has been chosen to make the spectral break near the lowest
observed velocities: $50^\circ$ for L1551, $40^\circ$ for NGC2071 and
NGC2264G. {\em Filled symbols}: blue lobes; {\em open symbols}: red
lobes. Velocities are relative to $v_0$, a center velocity, chosen to
maximize the symmetry between each pair of lobes.  Curves have been
offset vertically for clarity. \label{fig:outflows}}
\end{figure}

The protostellar winds described above are essentially radial, both
because they expand to fill the available angle, and because the wind
coasts after achieving force balance. At wind speeds typical of
low-mass protostars, the shocks that separate the wind from the
surrounding gas are radiative \markcite{KM92a}({Koo} \& {McKee} 1992),
so the swept-up shell conserves momentum in each direction. For a
steady wind and an ambient medium whose density varies as $\rho_a
\propto Q(\theta) r^{-2}$, \markcite{Shuea91}{Shu} {et~al.} (1991)
showed that these ingredients reproduce the Hubble law for
outflows. However, \markcite{1999ApJ...526L.109M}{Matzner} \& {McKee}
(1999a) demonstrated that neither a steady wind nor $\rho_a\propto
r^{-2}$ is necessary for this conclusion: as long as the ambient
medium is relatively featureless (e.g., a power law) on the scale of
the outflow, the outflow motion is self-similar even if the wind
intensity varies. For radial motion this implies the Hubble law, since
${\mathbf v}(\theta,t)\propto {\mathbf r}(\theta,t)/t$. 
The collimated wind force distribution $\rho_w v_w\propto
(\sin\theta)^{-2}$ leads to outflows that are highly elongated and
have bowshock-shaped tips of width $r\theta_0$. This morphology is
thus consistent with a ``wide-angle'' driving wind. A collimated
outflow with ${\mathbf v}\propto {\mathbf r}$ lacks blue material in
its red lobe and vice versa, consistent with observations.

That leaves only the power-law distribution of mass with velocity,
$dm/dv_{\rm obs}\propto v_{\rm obs}^{-1.8}$, to be
explained. \markcite{MC92}{Masson} \& {Chernin} (1992) argued that the
radial outflow model of \markcite{Shuea91}{Shu} {et~al.} (1991) was
inconsistent with this relation, but they only considered angular
distributions $\rho_w v_w^2\propto (\cos\theta)^{-\beta}$ for some
$\beta$. Their results do not apply to hydromagnetic winds, for which
$\rho_w v_w^2 \propto (\sin\theta)^{-2}$. Indeed,
\markcite{1996ApJ...472..211L}{Li} \& {Shu} (1996) demonstrated that
the outflow driven by a steady x-wind into a magnetically-flattened
core, with $\rho_a \propto Q(\theta) r^{-2}$, could produce
$dm/dv_{\rm obs}\propto v_{\rm obs}^{-2}$. I shall now show that this
conclusion is much more general
\markcite{1999ApJ...526L.109M}(see also {Matzner} \& {McKee} 1999a),
and is essentially independent of the ambient medium. 

The outflow shell expands at the rate dictated by momentum
conservation; since $\rho_w v_w^2 \propto \theta^{-2}$ for small
$\theta$, and since one expects the ambient medium also to be a power
law of $\theta$, the rate of expansion obeys a power-law relation
$v(\theta)\propto \theta^{-x}$ quite generally. The power $x$ can be
computed, but it is not important for the present argument. Now,
hydromagnetic winds satisfy the cumulative momentum distribution
$p_w(<\theta)\propto \ln(\theta/\theta_0)$ for a wide range of angles,
as described in \S 3. But since $v\propto \theta^{-x}$, momentum
conservation [$p_{\rm shell}(<\theta)=p_w(<\theta)$] implies a
cumulative momentum distribution with outflow velocity $p_{\rm
shell}(>v) \propto \ln(v/v_{\rm max})$, where $v_{\rm max}$ is the
rate of expansion along the wind axis. This, in turn, implies that
$dm/dv = d^2p_{\rm shell}(>v)/dv^2 \propto v^{-2}$; and, since $v_{\rm
obs} \propto v$ for an elongated outflow, $dm/dv_{\rm obs}\propto
v_{\rm obs}^{-2}$, i.e., $\Gamma =-2$. Notice that essentially nothing
about the ambient medium entered into this argument; therefore,
$\Gamma\simeq -2$ is generic.
Figure 1 demonstrates that models with this scaling can reproduce the
mass-velocity distributions of real outflows.

The commonly-observed features of molecular outflows are thus
the natural products of momentum-conserving shells driven by
hydromagnetic protostellar winds: they are insensitive to the details
of the wind's launching region, of variations in the wind's intensity,
and of the ambient density distribution. Although this model remains to be
tested in detail, it matches observation well enough to deserve
attention as a mechanism for feedback. 

\section{The efficiency of star and star cluster formation and the
IMF}\label{S:effic} The intensity of a protostellar wind along its
axis ensures that within some angle, gas will be blown away. Because
nearby gas is typically accreting onto the star itself or in the
process of forming other stars, this process limits the efficiency
with which individual stars and multiple stellar systems can form
\markcite{effic}(see {Matzner} \& {McKee} 2000, for a more detailed
discussion). Gas dispersal by protostellar winds is more ubiquitous
and less violent than disruption by massive stars.

To estimate the amount of gas ($\Mej$) lost per outflow, first note
that outflows propagate by conserving momentum in each direction. To
be ejected, material must travel faster than the escape velocity
$\vesc$ of the system. The momentum of the escaping gas is at least
$\Mej\vesc$; therefore, $\Mej\vesc < f_w v_w m_\star$. In reality,
momentum is lost in those directions where the wind is too weak to
eject anything, and wasted in those where the ejected gas travels
faster than $\vesc$. Since the wind momentum is distributed
logarithmically with angle, the fraction available to eject material
at around $\vesc$ is roughly $\ln(1/\theta_0)$; in fact,
$\ln(2/\theta_0)$ turns out to be a better estimate. Moreover, an
outflow's momentum is reduced by a factor $c_g$ (typically of order
unity) due to the action of gravity as it traverses the
cloud. Defining the {\em efficiency parameter} $X$ and the efficiency
per star $\varepsilon$,
\begin{equation}\label{eq:X}
X \equiv c_g \ln(2/\theta_0) \vesc/(f_w v_w), 
~~ \varepsilon \equiv m_\star/(m_\star+\Mej), 
\end{equation} 
{Matzner} \& {McKee} (2000) find that for the formation of an
individual star from a collapsing molecular core,
\begin{equation}\label{eq:coreeffic}
\varepsilon^{-1} \simeq (2X)^{-1} + [(2X)^{-2} + (1+f_w)^2]^{1/2}; 
\end{equation} 
and for the formation of a star within a larger cluster-forming clump, 
\begin{equation}\label{eq:effic}
\varepsilon^{-1} \simeq 1 + (2X)^{-1}. 
\end{equation} 

\begin{figure}
\plottwo{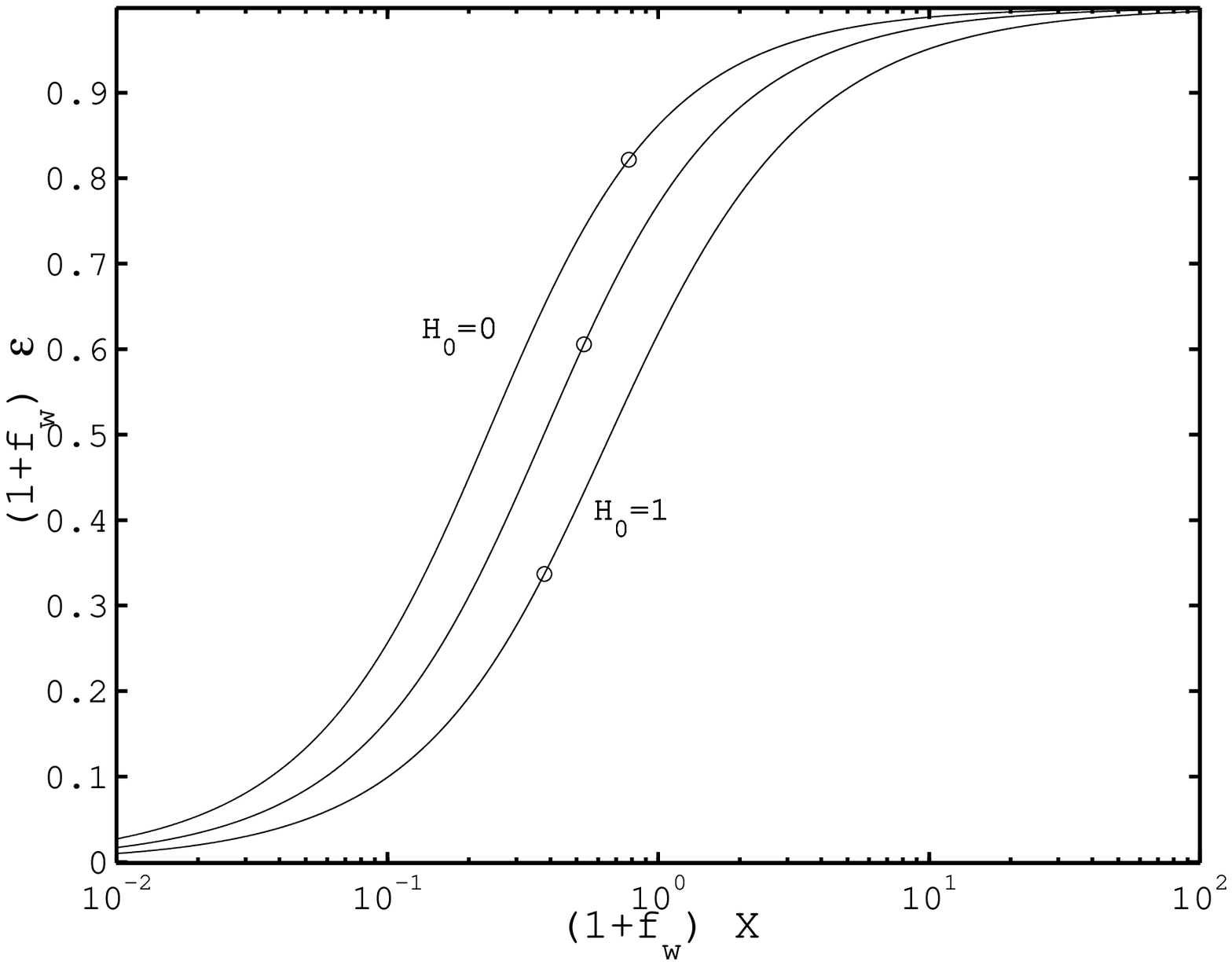}{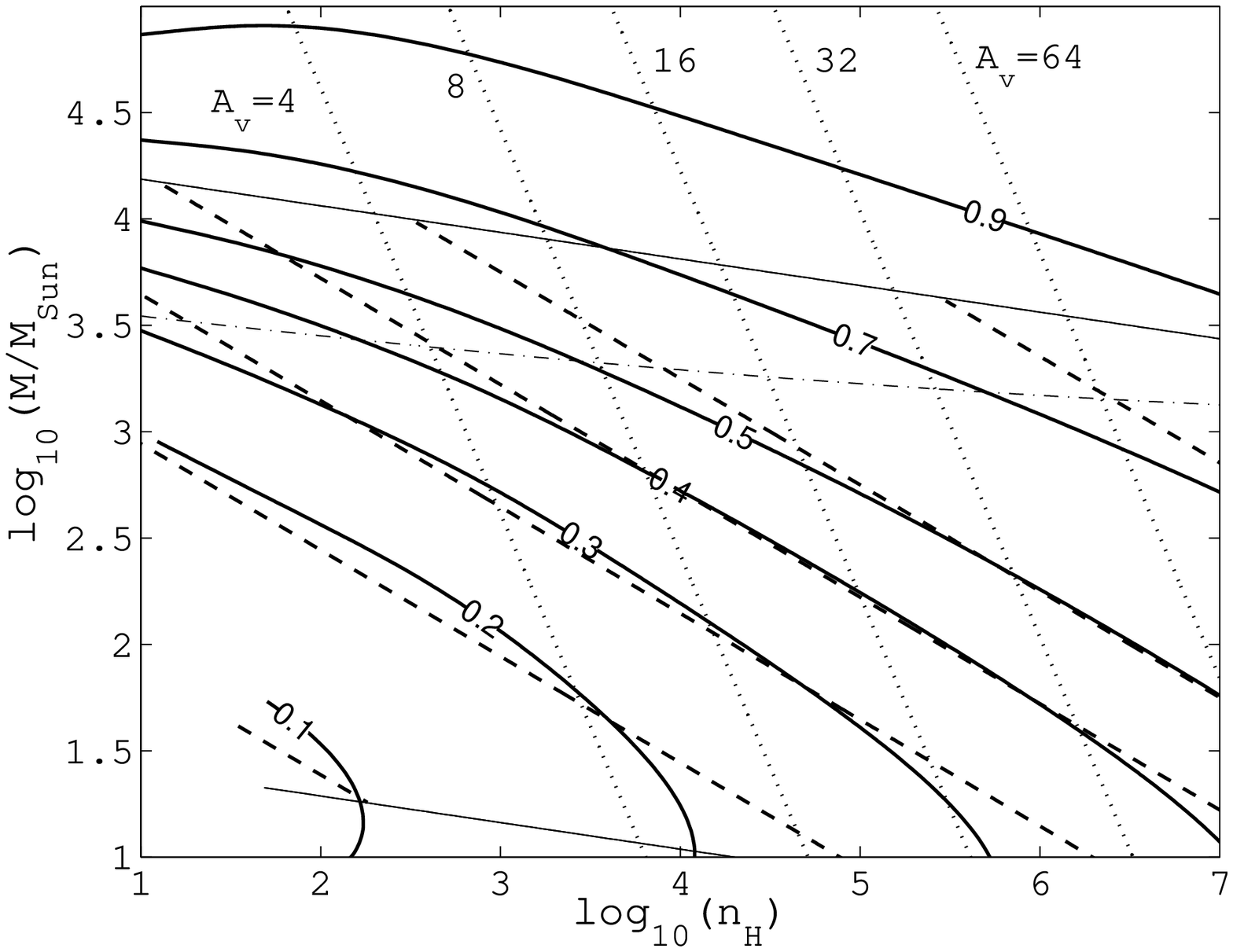}
\caption{\small {\em Left panel}: 
Core disruption and the efficiency of single star formation. The
efficiency $\varepsilon$ of protostellar core collapse (equation
\ref{eq:coreeffic}), versus the efficiency factor $X$. Curves are
plotted for spherical ($H_0=0$), moderately flattened ($H_0 = 0.4$)
and significantly flattened ($H_0=1$) cores. {\em Circles} indicate
the values of $X$ for thermally supported ($T = 10~{\rm K}$) cores;
turbulence can only increase $X$ above these values.
{\em Right panel}: 
Estimates of the instantaneous star formation efficiency $\varepsilon$
in stellar cluster formation. {\em Dashed lines:} analytical estimate,
equation (\ref{eq:effic}).  {\em Solid lines}: Numerical evaluation
of the same quantity in a protostellar clump with the structure
$\rho\propto r^{-1}$ and in which stars form at the local ambipolar
diffusion rate.  {\em Dotted lines} indicate the mean extinction of
the clump. Above the {\em dash-dot line}, stars more massive than
$20~\Msun$ are likely to form; our estimates are therefore only rough
upper limits in this region. The analytical expression relies on
assumptions that break down outside of the {\em upper and lower thin lines}. 
\label{fig:Core-Clump-Effic}}
\end{figure}
%

In both cases, $\varepsilon\propto X$ for $X\ll 1$ (strong winds) and
$\varepsilon \rightarrow 1$ for $X \gg 1$ (weak winds). The core and
clump cases differ for a number of reasons. In the formation of a single
star, unlike in multiple star formation, we can safely assume that all
of the core mass is either accreted or ejected. Clumps have higher
values of $\vesc$ (which raises $X$), but their density profiles are
shallower (which lowers $X$ through $c_g$); these effects roughly
cancel. Finally, mass is ejected at angles much further from the wind
axis in the core case; for this reason, the ejected wind 
mass is included in (\ref{eq:coreeffic}) but neglected in (\ref{eq:effic}). 

For the formation of an individual star, (\ref{eq:coreeffic}) assumes
that the collapsing core is spherically symmetric. However,
protostellar cores are likely to be partially supported by magnetic
fields, which will cause them to flatten along the magnetic
axis. Since the protostellar wind is likely to share this axis,
magnetic flattening segregates material away from the wind axis,
reducing the amount that will be ejected and increasing
the efficiency. Employing the models of flattened, magnetized cores
presented by \markcite{1996ApJ...472..211L}{Li} \& {Shu} (1996),
\markcite{effic}{Matzner} \& {McKee} (2000) find that equation
(\ref{eq:coreeffic}) remains valid if $X$ is replaced by
$(1+H_0)^{3/2} X$, where $1\leq (1+H_0)\lesssim 2$ is the factor by
which magnetic support increases the core's mean density. In the limit
of a completely disk-like core ($H_0\gg 1$), only the wind mass would
be lost [$\varepsilon \rightarrow 1/(1+f_w)$]. Individual stars form at about
$30\%$ efficiency if cores are spherical ($H_0=0$), or $\sim75\%$ if
they are significantly flattened ($H_0=1$), as shown in the left panel
of figure 2. 

For the formation of a star within a stellar cluster-forming clump,
the anisotropy of the region is not likely to correlate with the wind
axis. But, (\ref{eq:effic}) assumes that each star forms at the {\em
center} of a spherical distribution.  Comparing again simulations in
which stars form in a distributed manner throughout a cloud,
\markcite{effic}{Matzner} \& {McKee} (2000) find that (\ref{eq:effic})
is an excellent approximation nevertheless, as shown in the right
panel of figure 2. Equation (\ref{eq:effic}) predicts efficiencies
$30\% \lesssim \varepsilon \lesssim 50\%$ for the conditions typical
of low-mass stellar cluster-forming regions, like those in Orion B
\markcite{L92}({Lada} 1992): each star expels one to two times its own
mass from the clump. It is difficult to compare $\varepsilon$ with its
observational analogue, the ratio of stellar to total mass, but the
two quantities are similar in a few well-studied cluster-forming
regions \markcite{effic}({Matzner} \& {McKee} 2000).

Using the efficiency of individual star formation, we can estimate the
degree to which the stellar IMF differs from the mass function of
pre-stellar cores \markcite{1995ApJ...450..183N}({Nakano} {et~al.}
1995), a topic that has recently come under observational scrutiny
\markcite{1998A&A...336..150M}({Motte}, {Andr\'e}, \& {Neri}
1998). This comparison, discussed in detail by
\markcite{effic}{Matzner} \& {McKee} (2000), involves a consideration
of how much $X$, and thus $\varepsilon$, differs among cores of
different masses. We find that the variation of $\varepsilon$ is quite
subtle, and that the slope of the IMF is thus only slightly flatter
than the slope of the core mass function.
%

It should be noted that equations (\ref{eq:coreeffic}) and
(\ref{eq:effic}) predict efficiencies significantly higher than those
found by \markcite{1995ApJ...450..183N}{Nakano} {et~al.} (1995), and
that this is a direct result of the collimation of protostellar
winds. By this process (in contrast to the effect of a massive star),
mass is lost continuously as stars form. The efficiencies found above
are consistent with the formation of bound clusters
\markcite{1983ApJ...267L..97M}({Mathieu} 1983), which raises the
question of why so few actually do remain bound
\markcite{1991fesc.book....3L}({Lada} \& {Lada} 1991). This could be
due to a violent event associated with massive star formation;
alternatively, clusters could be unbound by motions of their parent
clumps, a topic we address below.

\section{Dynamical Evolution of Clumps During Cluster
Formation}\label{S:dynamics} 

Embedded stellar clusters are found exclusively within the most
massive clumps inside molecular clouds. Since these clumps are massive
enough to be confined by their own self-gravity despite the weight of
the molecular cloud above them \markcite{Crete99}({McKee} 1999), we
can gain insight into the production of a stellar cluster by
approximating a clump as a distinct reservoir of gas, and examining
its dynamical evolution while a stellar cluster forms within it. This
is the subject of an ongoing investigation
\markcite{myphd,1999sf99.proc..353M,MBM2000}({Matzner} 1999; {Matzner}
\& {McKee} 1999b; {Matzner} {et~al.} 2000), and the results reported
here should be considered preliminary.

Like any self-gravitating gaseous system, a cluster-forming clump must
satisfy the virial theorem for self-gravitating gas
(\markcite{1992ApJ...399..551M}here in Eulerian form; {McKee} \&
{Zweibel} 1992):
\begin{equation}\label{eq:Virial}
\frac{1}{2}\ddot{I} = 2({\cal T-T}_0)+{\cal W+M} -
\frac{1}{2}\frac{d}{dt}\int_S (\rho{\mathbf v}r^2)\cdot dS, 
\end{equation}
where ${I}\simeq M_{\rm cl} R_{\rm cl}^2/2$ is the trace of the
clumps's moment of inertia tensor; $\cal W$ and $\cal M$ are its
gravitational and magnetic energies; the kinetic energy of its
internal motions is $\cal T$; ${\cal T}_0$ is an energy associated
with its confining pressure; and the last term accounts for gas that
may leave the system. Equation (\ref{eq:Virial}) has been used to
identify equilibrium states of star-forming molecular clouds
\markcite{M89,BM96}({McKee} 1989; {Bertoldi} \& {McKee} 1996), but it
can also be taken as a dynamical equation if not taken to equal
zero. In order to accomplish this, each of the energetic terms must be
approximated as a function of four dynamical variables: the clump's
mass $M_{\rm cl}$ and the mass $M_\star$ of its embedded cluster, its
radius $R_{\rm cl}$, and its internal velocity dispersion $v_{\rm
rms}$. To simplify matters, the results presented here assume a
spherical, magnetically supercritical ($M_{\rm cl} = 2M_\Phi$) clump
with a $\rho\propto r^{-1}$ density profile and no rotation; any stars
are assumed to move along with the gas.

To complete our description of the system, it is necessary to specify
the evolution of $M_{\rm cl}$, $M_\star$, and $v_{\rm rms}$. All three
are affected by star formation and the effects of protostellar
outflows on the gas. For this, we use the theory of
\markcite{M89}{McKee} (1989): the local rate of gas conversion into
stars ($t_{g\star}^{-1}$) is identical to the local
ambipolar-diffusion rate ($t_{\rm AD}^{-1}$), because the formation of
protostellar cores requires the assembly of magnetically supercritical
condensations. Because of ionization by external FUV photons, star
formation is inhibited within four visual magnitudes of the surface;
interior to that, it proceeds at about a twelfth of the local
free-fall rate. As stars form, their mass is added to $M_\star$ and
subtracted from $M_{\rm cl}$; also subtracted is the mass of gas that
is ejected according to equation (\ref{eq:effic}).

For a massive molecular clump, $v_{\rm rms}$ is dominated by
non-thermal turbulent motions. The turbulent energy decays at a
characteristic rate that can be parameterized by $(d{v^2}_{\rm
rms}/dt)_{\rm decay} = -v_{\rm rms}^2/(\eta t_{\rm ff})$, where $t_{\rm ff}$
is the free-fall time of the clump. Therefore, turbulence must be
replenished every $\eta$ free-fall times; we expect $1\lesssim \eta
<10$, where numerical experiments \markcite{1998ApJ...508L..99S}(e.g.,
{Stone}, {Ostriker}, \& {Gammie} 1998) favor $\eta\lesssim 1$. Turbulent
energy can derive from two sources. For one, it can be exchanged for
other energies of the system, through work done during compressions or
expansions of the cloud; for this, we use the analytical results of
\markcite{1995ApJ...439..779Z}{Zweibel} \& {McKee} (1995) and
\markcite{1995ApJ...440..686M}{McKee} \& {Zweibel} (1995) to estimate
the energetics of turbulent, magnetized gas. This process does not,
however, prevent the cloud from collapsing. Alternatively, turbulence
may be stirred up by protostellar outflows. Since these conserve
momentum, it is most convenient to consider the clump's turbulent
momentum ($M_{\rm cl}v_{\rm rms}$) enhanced an amount $\phi_w p_w$ by
a wind of momentum $p_w$; here, $\phi_w\sim 0.6$ is the fraction of
the wind momentum that does not escape the clump
\markcite{myphd}({Matzner} 1999).

\begin{figure}\label{fig:etas}
\plotone{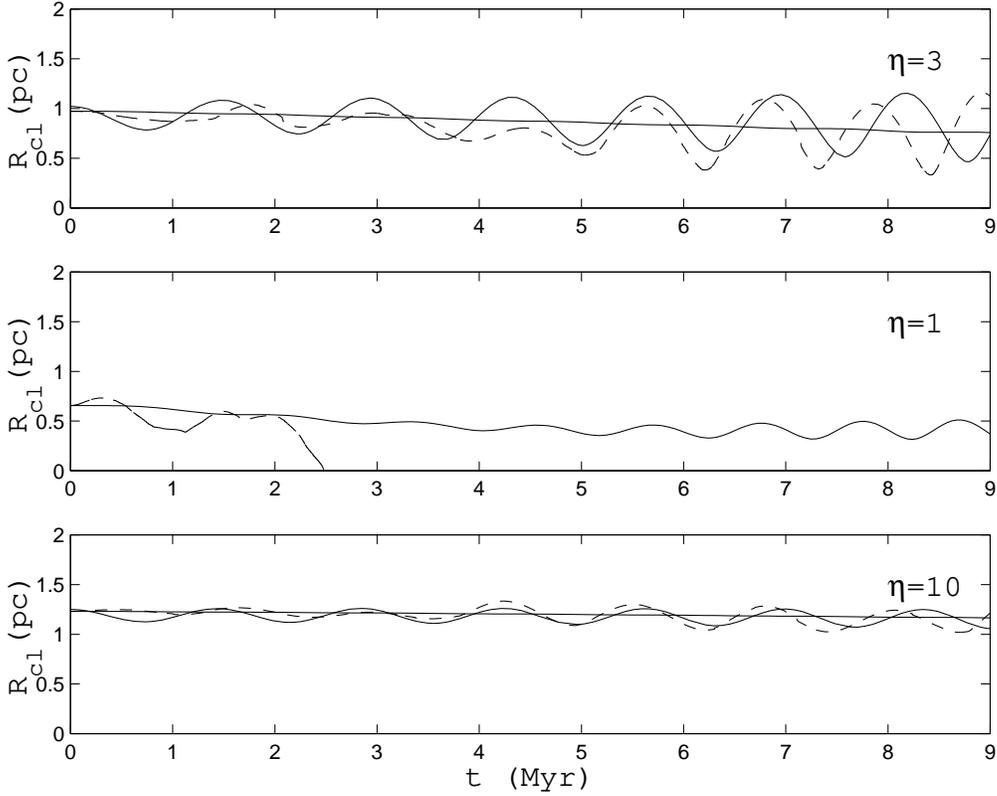}
\caption{\small The evolution of a $500~\Msun$ molecular clump as a stellar
cluster forms within it. All simulations start from a state of
equilibrium star formation; they differ in the rate of turbulent decay
$(\eta t_{\rm ff})^{-1}$, and in whether star formation is taken to be
continuous ({\em solid lines}) or individual ({\em dashed lines}).}
\end{figure}
Putting these elements together, it is useful to consider two
approximations: one in which energy is added continuously according to
the mean star formation rate within the clump, and another where stars
form randomly at the current star formation rate, are chosen
individually from the IMF, and affect the cloud instantaneously. These
approaches differ for clumps of about five hundred solar masses or
less, in which only ten to twenty stars form per $t_{\rm ff}$. In the
continuous approximation it is possible to identify equilibria
\markcite{NS80,M89}({Norman} \& {Silk} 1980; {McKee} 1989) in which
turbulent decay is just offset by injection from outflows. These are
stable, in the sense that a compressed cloud's turbulence is enhanced
faster than it decays. But they can also be overstable, as energy in
an oscillation can grow during compressive phases.

Perturbations due to individual forming stars will stimulate
oscillations of the system, whose response depends critically on the
rate of turbulent decay.  Figure 3 shows the effect of changing $\eta$
for a clump of $500M_\odot$: when turbulence decays slowly ($\eta =
10$), a cluster-forming cloud will remain very close to its
equilibrium state, whether or not star formation is taken to be
continuous. For moderate turbulent decay ($\eta=3$), individual star
formation seeds overstable oscillations about equilibrium: stars
form in bursts. For fast turbulent decay ($\eta=1$), a continuous
simulation can remain near its equilibrium, but one with individual
star formation cannot consistently replenish the lost turbulent energy
every dynamical time; the system collapses. Although the
approximations used in these simulations (e.g., spherical symmetry and
no rotation) affect these outcomes, it appears that the duration 
and constancy of star formation are sensitive to the rate of turbulent
decay. 

\section{Conclusions} 
Observations of star-forming regions and of individual accreting
protostars have made it increasingly clear that stars interact
violently with their parent clouds as they form. To model this
phenomenon requires an understanding of protostellar winds, and of the
bipolar molecular outflows they drive into the surrounding gas. We
have seen that the typically observed features of molecular outflows
can be explained as the natural product of magnetically-collimated
winds and radiative shocks. This yields a model for outflows which
can be used to estimate how much mass is ejected, either from an
individual collapsing core or from a massive clump in the process of
forming a stellar cluster. The estimates of star formation
efficiency ($30\% - 50 \%$) presented in \S 5 are much
higher than those of Nakano et al. (1995), because they account for
the collimation of protostellar winds. 

Lastly, the feedback effects of protostellar outflows can be
incorporated into simple dynamical models for the process of stellar
cluster formation within massive molecular clumps. The investigations
described in \S 6 show that the dynamics of stellar
cluster formation are sensitive to the rate at which turbulence
decays, and this will have observational consequences. 

In the future, large-scale numerical simulations of star formation
must include protostellar outflows, for without them such simulations
are incomplete. 

\acknowledgements I am grateful to Lee Hartmann, Zhi-Yun Li, Scott
Kenyon, Chris McKee and Frank Shu for many useful and insightful
comments. Motivation for the models of \S 6 came from a conversation
with Frank Bertoldi. I very much appreciate the gracious hospitality
of Roger Blandford and Sterl Phinney during my visits to Caltech, and
of Lars Bildsten during a visit to the ITP. This research was
supported in part by the National Science Foundation through NSF
grants AST 95-30480 and PHY 94-07194, in part by a NASA grant to the
Center for Star Formation Studies, and in part by an NSERC fellowship.


\end{document}